\documentclass[12pt]{iopart}
\usepackage{iopams}
\usepackage{bm}
\usepackage{dsfont}
\usepackage{graphicx}
\usepackage{setstack}
\usepackage{mathrsfs} 
\usepackage{xcolor}
\usepackage{braket}
\usepackage[colorlinks=true,breaklinks=false,allcolors=blue]{hyperref}

\usepackage{etoolbox}

\usepackage{lipsum}

\usepackage{bbm}
\usepackage{enumitem}

\definecolor{michael}{rgb}{0,.8,.5}

\newcommand\ZZ{{\mathbbm{Z}}}
\newcommand\RR{{\mathbbm{R}}}

\newcommand\one{{\mathbbm{1}}}

\newcommand\dd{{\mathrm{d}}}
\newcommand\ee{{\mathrm{e}}}
\newcommand\ii{{\mathrm{i}}}

\newcommand\dist{{\mathit{d}}}
\newcommand\Cor{{\mbox{Cor}}}

\begin{document}

\title{Entanglement-enhanced spreading of correlations}  

\author{Michael Kastner} 
\address{National Institute for Theoretical Physics (NITheP), Stellenbosch 7600, South Africa} 
\address{Institute of Theoretical Physics,  University of Stellenbosch, Stellenbosch 7600, South Africa}
\ead{kastner@sun.ac.za} 

\date{\today}

\begin{abstract}
Starting from a product initial state, equal-time correlations in nonrelativistic quantum lattice models propagate within a lightcone-like causal region. The presence of entanglement in the initial state can modify this behaviour, enhancing and accelerating the growth of correlations. In this paper we give a quantitative description, in the form of Lieb-Robinson-type bounds on equal-time correlation functions, of the interplay of dynamics {\em vs}.\ initial entanglement in quantum lattice models out of equilibrium. The bounds are tested against model calculations, and applications to quantum quenches, quantum channels, and Kondo physics are discussed.
\end{abstract}



\section{Introduction}

Correlations are a quantity of great importance in statistical and condensed matter physics, and they represent an essential resource in quantum information science. In traditional condensed matter systems, correlations are often measured in scattering experiments. More recently, technological advances have established trapped ultracold atoms and ions as versatile experimental platforms for the study of many-body quantum systems. Owing to the high level of precision and control in such experiments, equal-time correlation functions can be measured at atomic spatial resolution and, simultaneously, with a temporal resolution much higher than the intrinsic dynamical time scales \cite{Cheneau_etal12,Richerme_etal14,Jurcevic_etal14}.

\begin{figure}\centering
\includegraphics[height=0.37\linewidth]{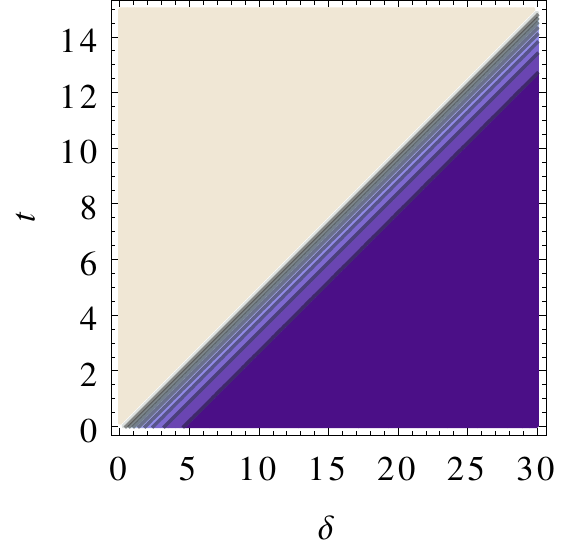}
\includegraphics[height=0.37\linewidth]{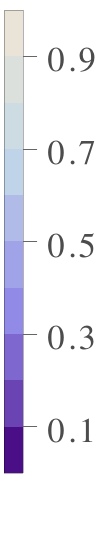}
\includegraphics[height=0.38\linewidth]{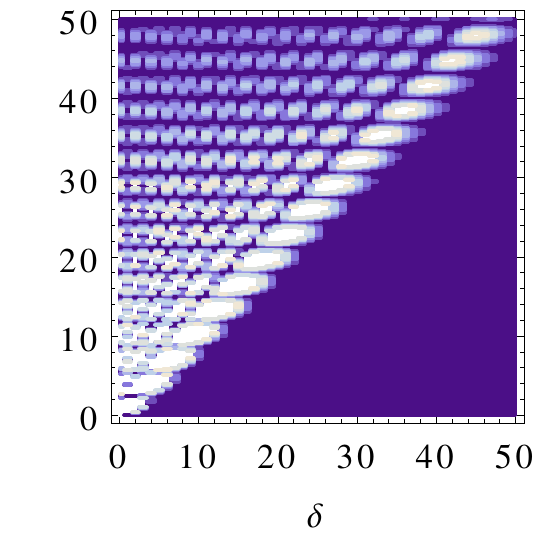}
\includegraphics[height=0.38\linewidth]{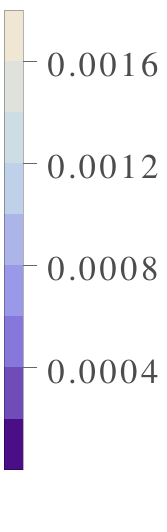}
\caption{\label{f:cone}%
Left: Lieb-Robinson bound for equal-time correlation functions between sites $i$ and $j$, plotted as a function of time $t$ and distance $\delta=d(i,j)$ between sites. Outside the light-coloured cone-like region, correlations are exponentially suppressed. Right: Exact analytic results for the absolute value of the connected correlations $\langle\sigma_i^x\sigma_j^x\rangle_\mathrm{c}$ in an $XX$ spin chain with nearest-neighbour interactions \eref{e:XX}, starting from the product initial state \eref{e:ProdIni}. Quasilocal behaviour, i.e., spreading with only exponentially small effects outside a cone-shaped region, is observed.
}
\end{figure}

Theoretically, the creation and propagation of correlations is well understood in the case of uncorrelated initial states, and also for exponentially clustered ones. Making use of Lieb-Robinson bounds \cite{LiebRobinson72}, rigorous estimates of the spatial and temporal behaviour of equal-time correlation functions have been derived for short-range interacting systems with exponentially clustered initial states \cite{BravyiHastingsVerstraete06}, and also for rather general types of interactions and uncorrelated initial states \cite{NachtergaeleOgataSims06}. The picture that emerges from these results is that of {\em quasilocality}. In the case of short-range interactions this means that correlations are approximately (up to exponentially small corrections) confined to a causal region, resembling the lightcone of a relativistic theory; see figure \ref{f:cone} (left) for an illustration.

\begin{figure}\centering
\includegraphics[width=0.47\linewidth]{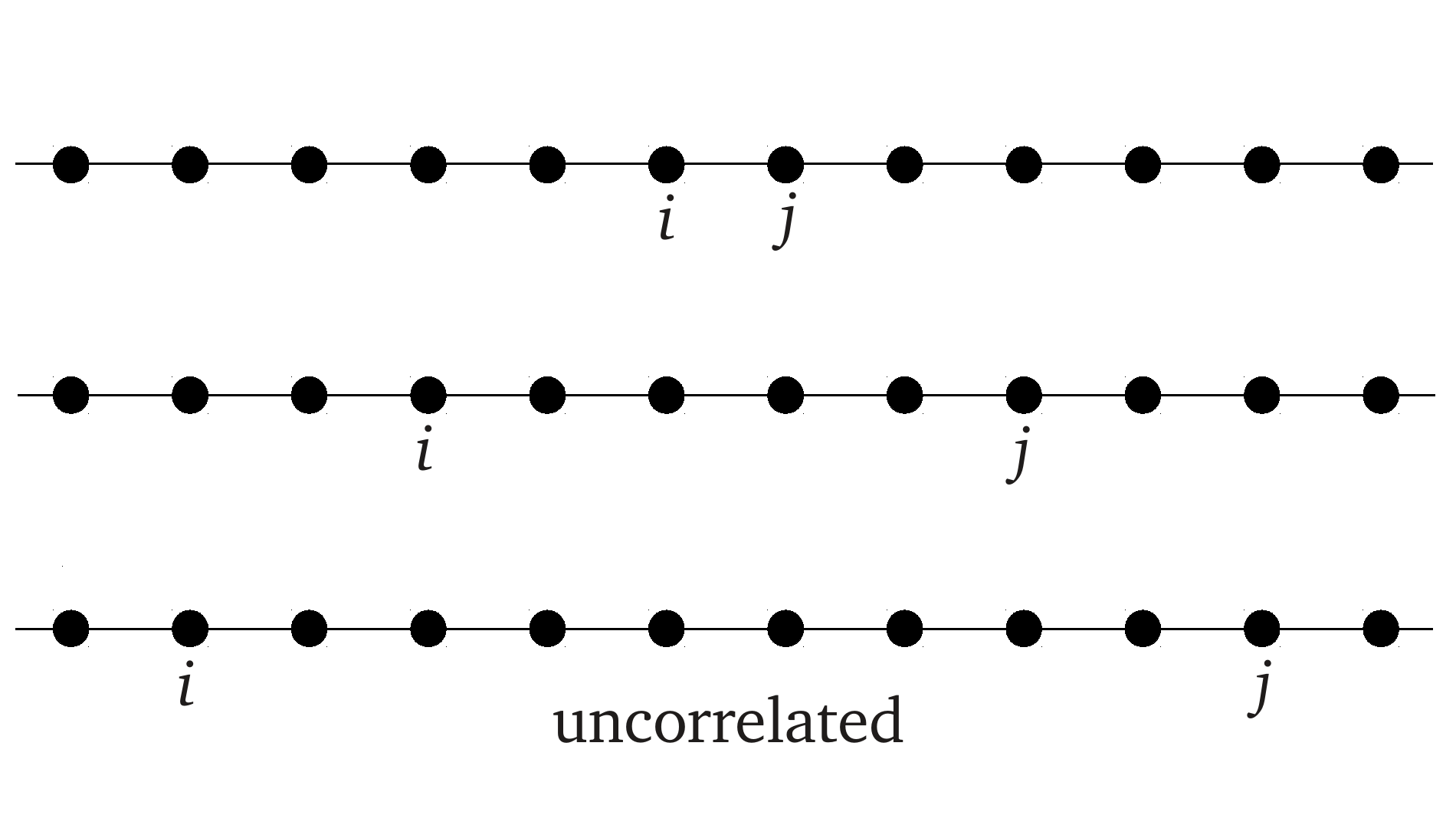}
\hfill
\includegraphics[width=0.47\linewidth]{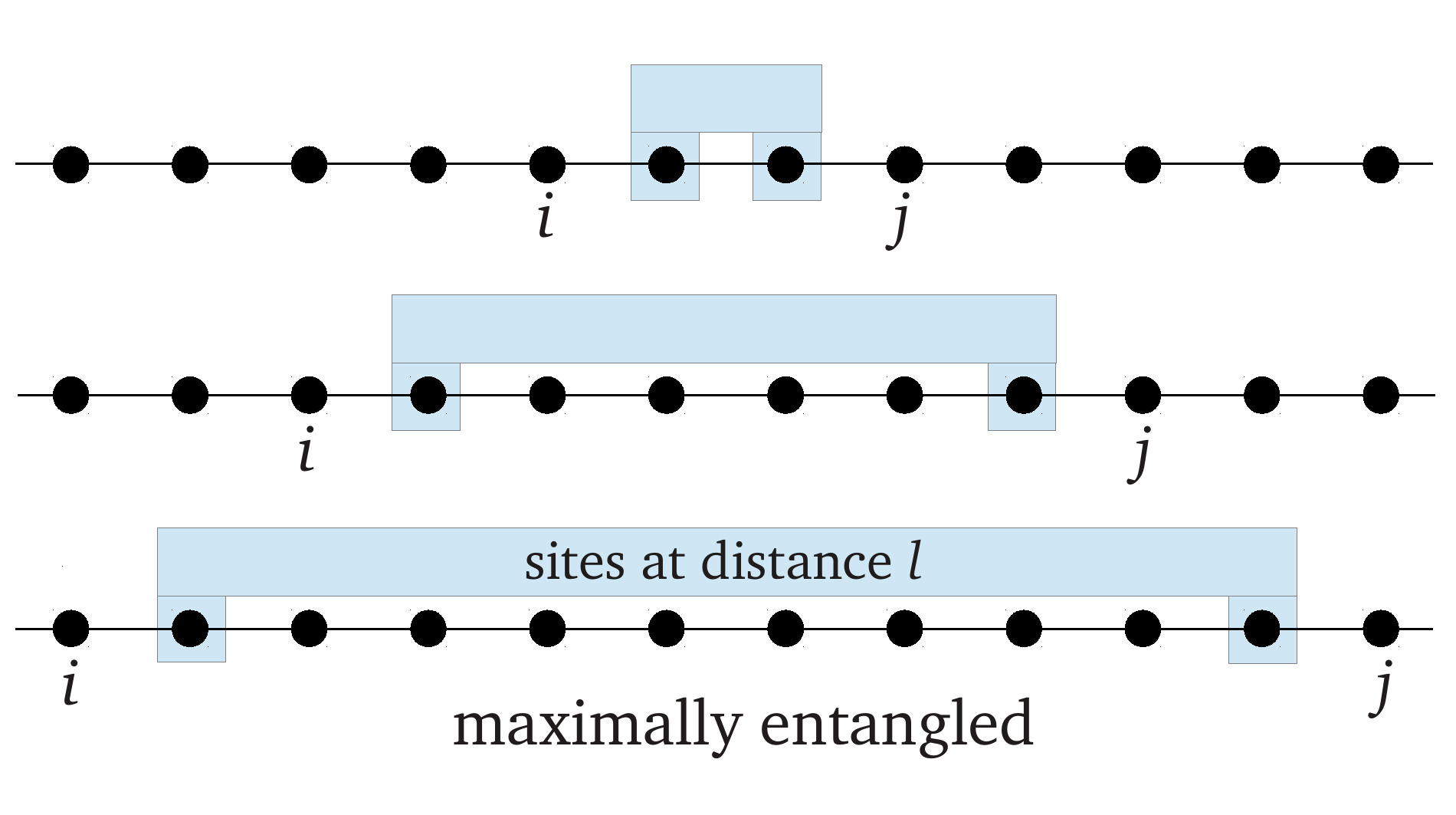}
\includegraphics[width=0.49\linewidth]{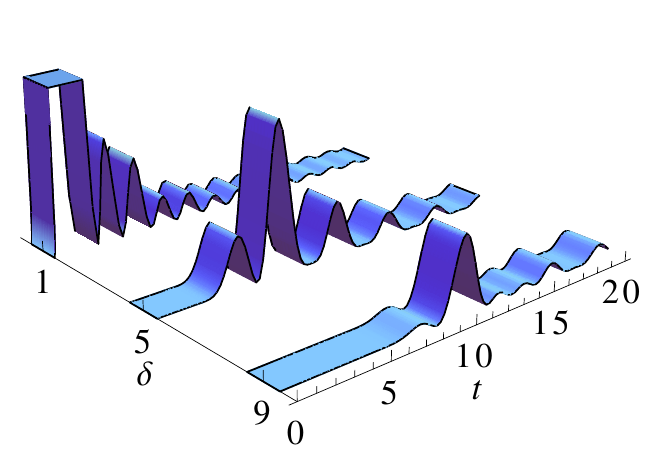}
\hfill
\includegraphics[width=0.49\linewidth]{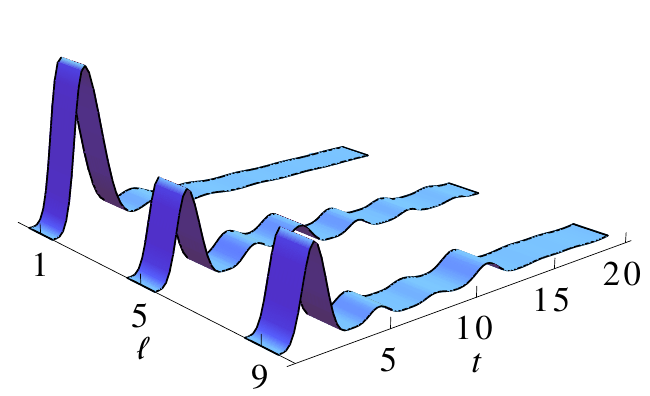}
\caption{\label{f:XX_Corr}%
Exact analytic results for connected correlation functions $\langle\sigma_i^x\sigma_j^x\rangle_\mathrm{c}$ of the $XX$-chain with nearest-neighbour interactions \eref{e:XX}. Left: Starting from a product initial state \eref{e:ProdIni}, correlations build up after a time $t$ that scales linearly with the distance $\delta$ between lattice sites $i$ and $j$. Right: For an entangled initial state where initial correlations extend over a distance $\ell=\delta-2$. In this case, correlations between sites $i$ and $j$ are created on a timescale that is independent of the distance $\delta$ between the sites.}
\end{figure}

As an example consider a spin chain with nearest-neighbour interactions, as sketched in figure \ref{f:XX_Corr} (top left). We are interested in the time evolution of connected spin--spin correlations
\begin{equation}\label{e:ConnCorr}
\langle\sigma_i^x\sigma_j^x\rangle_\mathrm{c} := \langle\sigma_i^x\sigma_j^x\rangle - \langle\sigma_i^x\rangle\langle\sigma_j^x\rangle
\end{equation}
between lattice sites $i$ and $j$, where $\sigma_i^x$ denotes the $x$-Pauli matrix at site $i$. Starting from a product state
\begin{equation}\label{e:ProdIni}
|\psi\rangle=|\downarrow\rangle_i\bigotimes_{k\neq i}|\uparrow\rangle_k
\end{equation}
where $\vert\uparrow\rangle_k$ denotes an eigenstate of $\sigma_k^z$, all connected correlations vanish initially. Under the time evolution induced by a Hamiltonian with nearest-neighbour interactions, correlations build up and spread in a distance-dependent fashion, as illustrated in figure \ref{f:cone} (right) and figure \ref{f:XX_Corr} (bottom left). The larger the distance $\delta$ between $i$ and $j$, the longer it takes for correlations between the sites to build up.

It should not come as a surprise that the presence of entanglement in the initial state can modify this picture. In the same way that performing a measurement on one constituent of a Bell pair has an instantaneous effect on its distant partner, so can interactions with entangled entities speed up the propagation of correlations. An illustration of this kind of behaviour is given in figure \ref{f:XX_Corr} (right). We consider again connected correlation functions $\langle\sigma_i^x\sigma_j^x\rangle_\mathrm{c}$ of a spin chain with nearest-neighbour interactions, this time starting from an initial state
\begin{equation}\label{e:EntIni}
|\psi\rangle=\left(|\uparrow\rangle_{i+1}|\downarrow\rangle_{j-1}+|\downarrow\rangle_{i+1}|\uparrow\rangle_{j-1}\right)\bigotimes_{k\neq i+1,j-1}|\uparrow\rangle_k
\end{equation}
that is mostly of a product form, with the exception of sites $i+1$ and $j-1$ being maximally (Bell) entangled. Due to the almost-product structure, correlations between sites $i$ and $j$ are initially vanishing. Under time evolution, the propagation of correlations is enhanced by the long-distance entanglement present in the initial state. For the specific, and rather artificially constructed, initial state \eref{e:EntIni}, correlations build up in a distance-independent fashion (figure \ref{f:XX_Corr} bottom right).

In general, and in particular for physically realistic initial states, the creation of correlations will be determined by an interplay of dynamical effects due to interactions on the one side, and of initial entanglement on the other side. In this paper we develop theoretical tools for a quantitative description of the propagation of correlations in the presence of initial entanglement. The main result \eref{e:general_final}--\eref{e:BoundFinal} is an upper bound on the connected correlation function, containing a Lieb-Robinson-type contribution capturing the dynamics, and a second term that takes into account initial correlations. Depending on the amount and shape of initial correlations, the bound is able to capture the extremes of product initial states (as in figure \ref{f:XX_Corr} left) on the one side and long-distance entangled initial states (as in figure \ref{f:XX_Corr} right) on the other side, as well as the more involved cases in between where the interplay of dynamics and initial entanglement 
leads to nontrivial propagation patterns.

The results of this paper apply to a broad class of quantum mechanical lattice models (detailed in section \ref{s:LR}) and arbitrary initial states, and this generality accounts for many potential physical applications. In fact, unless specifically prepared, a product initial state should be considered the exception rather than the rule. Physical applications in which long-distance correlated initial states play an important role include:
\begin{enumerate}[label={(\alph*)}]
\item\label{a} Quenching away from a quantum critical point. A simple, exactly solvable example is a spin-$1/2$ Ising chain in a transverse magnetic field of strength $h$ \cite{Pfeuty70}. At a critical value $h=h_\mathrm{c}$ of the field strength the model undergoes a quantum phase transition from a ferromagnetic to a paramagnetic phase. Preparing the system at $h_\mathrm{c}$ in the ground state, connected correlation functions between spins at sites $i$ and $j$ decay like a power law with the distance between the sites. A quench, i.e., a sudden change of the Hamiltonian, then triggers a time evolution, and the propagation of a local perturbation will be affected by the presence of long-distance initial correlations.
\item\label{b} Quantum transport and qubit transfer in spin chains with long-distance entangled ground states. Examples are dimerised open chains, some of which are known to have ground states with long-distance entanglement between the end points of the chain \cite{Campos_etal06}. When such a chain is used as a quantum channel, entanglement-enhanced propagation is observed \cite{Campos_etal07}, reminiscent of the scenario depicted in figure \ref{f:XX_Corr} (right).
\item\label{c} Building-up of a Kondo screening cloud in the vicinity of an impurity spin. In the Kondo model, an impurity spin is coupled to a noninteracting Fermi gas. At zero temperature the spatial correlations in the Fermi gas decay like a power law with the distance \cite{FetterWalecka}. Starting from an initial state with no correlations between impurity and Fermi gas, analytical \cite{MedvedyevaHoffmannKehrein13} as well as numerical \cite{LechtenbergAnders14} calculations show that correlations build up predominantly in a cone-shaped region in space-time, but with a slow (power law) spatial decay outside the cone. \end{enumerate}
In section \ref{s:LR} a Lieb-Robinson bound in a rather general setting is reviewed. In sections \ref{s:comparing}--\ref{s:correlations} our main result, a bound on the connected correlation function in the presence of long-distance initial entanglement is derived. Specific types of interactions and initial entanglement distributions are discussed in sections \ref{s:specific} and \ref{s:examples}. We find good qualitative agreement when comparing our bounds to model calculations, and also discuss the bounds in the context of examples \ref{a}--\ref{c}.

\section{Lieb-Robinson bounds}
\label{s:LR}

The strategy is to use Lieb-Robinson bounds to construct an upper bound on equal-time correlation functions. As a setting we choose the rather general class of quantum lattice models of \cite{NachtergaeleOgataSims06} for which Lieb-Robinson bounds, and also bounds on equal-time correlation functions in the case of product initial states, have been derived. The following notation and conditions are similar to those in \cite{NachtergaeleOgataSims06}, but we have simplified the presentation, and in particular avoided the $C^*$-algebraic language used in that reference. 

On a graph $\Lambda$ we have a finite-dimensional Hilbert space $\mathscr{H}_i$ at each vertex $i\in\Lambda$. On the tensor product space $\mathscr{H}_\Lambda=\bigotimes_{i\in\Lambda}\mathscr{H}_i$ the Hamiltonian
\begin{equation}
H:=\sum_{X\in\Lambda}\Phi(X)
\end{equation}
is defined, where the interaction $\Phi(X)$ is a bounded linear operator acting nontrivially only on the part of the Hilbert space that is associated with the subset $X\subset\Lambda$. We denote the time evolution of a bounded linear operator $A$ in the Heisenberg picture as
\begin{equation}
A(t)=\ee^{\ii H t}A\ee^{-\ii H t}.
\end{equation}

To be able to prove a Lieb-Robinson bound, the interactions $\Phi(X)$ need to decay with the spatial separation of the lattice sites in $X$ in a suitable way. This is enforced by requiring
\begin{equation}
\|\Phi\|:=\sup_{i,j\in\Lambda}\sum_{X\ni i,j}\frac{\|\Phi(X)\|}{F(\dist(i,j))}<\infty,
\end{equation}
where $F:[0,\infty)\to(0,\infty)$ is a positive function characterizing the spatial decay of the interactions, and $\dist$ is the graph distance on $\Lambda$. The requirements on the spatial decay function are that $F$ is uniformly summable over $\Lambda$,
\begin{equation}\label{e:integrability}
\| F\|:=\sup_{i\in\Lambda}\sum_{j\in\Lambda}F(\dist(i,j))<\infty,
\end{equation}
and that it satisfies
\begin{equation}\label{e:reproducibility}
C:=\sup_{i,j\in\Lambda}\sum_{k\in\Lambda}\frac{F(\dist(i,k))F(\dist(k,j))}{F(\dist(i,j))}<\infty.
\end{equation}
On a regular lattice like $\Lambda=\ZZ^D$, equations \eref{e:integrability} and \eref{e:reproducibility} are satisfied for example by $F(x)\propto(1+x)^{-\alpha}$ for $\alpha>D$, which is a suitable choice for pair interactions decaying asymptotically for large distances according to a power law with exponent $-\alpha$. For interactions of finite range (like nearest-neighbour interactions), or for exponentially decaying interactions, $F(x)\propto\ee^{-ax}/(1+x)^{D+1}$ with $a>0$ is a suitable choice \footnote{As pointed out in \cite{NachtergaeleOgataSims06}, a pure exponential $F(x)=\ee^{-ax}$ does not satisfy \eref{e:reproducibility}.}.

Considering bounded linear observables $A$, $B$ acting nontrivially only on the regions $X,Y\subset\Lambda$, respectively, a Lieb-Robinson bound
\begin{equation}\label{e:LR}
\left\|\left[A(t),B\right]\right\|\leq\frac{2\| A\| \| B\|}{C}g(t)\sum_{i\in X}\sum_{j\in Y}F(\dist(i,j))
\end{equation}
holds for any $t\in\RR$, where
\begin{equation}\label{e:g}
g(t)=\cases{
\exp(2\|\Phi\| C|t|)-1 & for $\dist(X,Y)>0$,\\
\exp(2\|\Phi\| C|t|) & else.
}
\end{equation}
For a proof, see section 2.1 of \cite{NachtergaeleOgataSims06}.

\section{Comparing time evolutions}
\label{s:comparing}

The bound \eref{e:LR} can be used to quantify the difference between an observable $A(t)$ that is time-evolved with the full Hamiltonian $H$, and an observable $A'(t)$ obtained by time-evolving $A$ with a modified Hamiltonian $H'$. To this aim, one can write
\begin{eqnarray}\label{e:truncdyn}
\| A(t)-A'(t)\| &=& \left\|\ee^{\ii H t}A\ee^{-\ii H t} - \ee^{\ii H't}A\ee^{-\ii H' t}\right\|\nonumber\\
&=& \left\|\ee^{-\ii H t}\ee^{\ii H' t}A\ee^{-\ii H' t}\ee^{\ii H t} - A\right\|\nonumber\\
&=& \left\|\int_0^t \dd\tau \frac{\dd}{\dd\tau}\ee^{-\ii H \tau}\ee^{\ii H' \tau}A\ee^{-\ii H' \tau}\ee^{\ii H\tau}\right\|\nonumber\\
&=& \left\|\int_0^t \dd\tau \ee^{-\ii H \tau}\left[H-H',A'(\tau)\right]\ee^{\ii H\tau}\right\|\nonumber\\
&\leq& \int_0^{|t|} \dd\tau \left\|\left[A'(\tau),H-H'\right]\right\|,
\end{eqnarray}
where unitarity of the time evolution operators and the triangle inequality were used (see equation (S10) of \cite{Gong_etal_arXiv} or Lemma 3.3 of \cite{NachtergaeleOgataSims06}). The integrand in the last line of equation \eref{e:truncdyn} has the form of the commutator on the left-hand side of the Lieb-Robinson bound \eref{e:LR}, which we can therefore use to further estimate \eref{e:truncdyn}.

\section{Decoupled dynamics}

In section \ref{s:comparing} we did not specify the way in which the Hamiltonian is modified, i.e., how $H$ and $H'$ are related. Since Lieb-Robinson bounds establish quasilocality of the time evolution, they can give a particularly useful estimate when comparing the full dynamics under a Hamiltonian $H$ with that of a ``decoupled'' Hamiltonian $H'$ in which all interactions between two spatial regions have been eliminated. Quasilocality then suggest considering a ball 
\begin{equation}
S_X:=\left\{j\in\Lambda\,:\,\dist(j,X)\leq r\right\}
\end{equation}
around the support $X$ of the observable $A$ as one spatial region, and the outside of the ball as the other one, and choosing a ball with radius $r$ large enough that the effect of the eliminated interaction terms on the time-evolved observable $A$ is small. The decoupled Hamiltonian is then given by
\begin{equation}
H'=\sum_{Z\in\Lambda:Z\cap S_X=\emptyset\;\mathrm{or}\;Z\cap S_X^\mathrm{c}=\emptyset}\Phi(Z),
\end{equation}
where $S_X^\mathrm{c}$ denotes the complement of $S_X$ with respect to $\Lambda$. $H-H'$ then contains all terms (and only those) that couple $S_X$ to its complement.

For these choices of $H$ and $H'$, and using the triangle inequality, we can bound the integrand in the last line of \eref{e:truncdyn} by
\begin{equation}
\left\|\left[A'(\tau),H-H'\right]\right\| \leq \!\sum_{Z\in\Lambda:Z\cap S_X\neq\emptyset\;\mathrm{or}\;Z\cap S_X^\mathrm{c}\neq\emptyset}\! \left\|\left[A'(\tau),\Phi(Z)\right]\right\|.
\end{equation}
Applying the Lieb-Robinson bound \eref{e:LR} to each of the terms in the sum, and following equations (3.10)--(3.15) of \cite{NachtergaeleOgataSims06}, one arrives at
\begin{equation}
\left\|\left[A'(\tau),H-H'\right]\right\| \leq 2g(\tau)\| A\|\|\Phi\|\frac{C+\| F\|}{C}\sum_{i\in X}\sum_{j\in S_X^\mathrm{c}}F(\dist(i,j)). 
\end{equation}
Inserting this expression into \eref{e:truncdyn}, we obtain a bound on $\| A(t)-A'(t)\|$, i.e., a bound on the size of the truncation error when comparing the time evolution of $A$ under the decoupled Hamiltonian $H'$ to the full time evolution under $H$. An analogous result is obtained for the time evolution of $B$ by using a ball $S_Y$ centred around the support $Y$ of $B$.


\section{Spreading of equal-time correlations}
\label{s:correlations}

Our main goal is to estimate equal-time connected correlation functions
\begin{equation}
\langle A(t)B(t)\rangle_\mathrm{c}:=\langle A(t)B(t)\rangle-\langle A(t)\rangle\langle B(t)\rangle,
\end{equation}
where $\langle\cdot\rangle=\Tr(\cdot\rho)$ denotes the quantum mechanical expectation value with respect to some initial state $\rho$. The strategy is to express the occurring operators in terms of differences $A(t)-A'(t)$, whose absolute value can be estimated by equations \eref{e:LR} and \eref{e:truncdyn}. To achieve this, similar to equation (S9) in \cite{Gong_etal_arXiv} we write
\begin{eqnarray}
\langle AB\rangle_\mathrm{c}&=&\langle AB\rangle+\underbrace{\langle A'B\rangle-\langle A'B\rangle}_{\displaystyle=0}+\underbrace{\langle A'B'\rangle-\langle A'B'\rangle}_{\displaystyle=0}\nonumber\\
&&\quad\;-\underbrace{\left(\langle A-A'\rangle+\langle A'\rangle\right)\left(\langle B-B'\rangle+\langle B'\rangle\right)}_{\displaystyle=\langle A\rangle\langle B\rangle}\nonumber\\
&=& \langle(A-A')B\rangle+\langle A'(B-B')\rangle-\langle A-A'\rangle\langle B'\rangle\nonumber\\
&&\quad\;-\langle A\rangle\langle B-B'\rangle + \langle A'B'\rangle-\langle A'\rangle\langle B'\rangle,
\end{eqnarray}
where the time dependences of the operators have been suppressed. A bound on the absolute value of the connected correlator is then given by
\begin{equation}\label{e:split}
\left|\langle AB\rangle_\mathrm{c}\right|\leq2\| A-A'\|\| B\|+2\| A\|\| B-B'\| + \left|\langle A'B'\rangle_\mathrm{c}\right|.
\end{equation}

In the case of a product initial state, $\left|\langle A'B'\rangle_\mathrm{c}\right|$ is zero as long as the radii $r$ of the balls $S_X$ and $S_Y$ centred around $X$ and $Y$, respectively, are non-overlapping, i.e., for $r\leq\dist(X,Y)$. In this case one recovers the result of \cite{NachtergaeleOgataSims06}. A related result in \cite{BravyiHastingsVerstraete06} permits exponentially (in space) decaying initial correlations, showing that, in the case of finite-range interactions, correlations are restricted to a cone-like region in space-time, with only exponentially small corrections outside the causal cone.

Here we want to allow for arbitrary initial correlations and investigate their effect on the creation of correlations in time. The fact that $A'$ and $B'$ evolve under decoupled dynamics imposes a restriction on the size of their correlations, which can be seen as follows. Divide the total Hilbert space into three factors,
\begin{equation}
\mathscr{H}_\Lambda = \mathscr{H}_{S_X} \otimes \mathscr{H}_{S_Y} \otimes \mathscr{H}_{\Lambda\setminus(S_X\cap S_Y)},
\end{equation}  
corresponding to different parts of the lattice as indicated by their indices. Since the decoupled dynamics does not mix between the factors of this tensor product, we can write
\begin{eqnarray}
\left|\langle A'B'\rangle_\mathrm{c}\right| &=& \left|\left\langle A'\otimes B'\otimes\one\right\rangle-\left\langle A'\otimes\one\otimes\one\right\rangle\left\langle B'\otimes\one\otimes\one\right\rangle\right|\nonumber\\
&\leq&\| A\|\| B\| \Cor(S_X:S_Y),
\end{eqnarray}
with
\begin{equation}\label{e:Cor}
\Cor(S_X:S_Y):=\max_{\Vert O_{S_X}\|,\| O_{S_Y}\|\leq1}\left|\langle O_{S_X}O_{S_Y}\rangle_\mathrm{c}\right|,
\end{equation}
where $O_{S_X}$ and $O_{S_Y}$ are observables supported on $S_X$ and $S_Y$, respectively. $\Cor(S_X:S_Y)$ quantifies, for a given state $\rho$ with respect to which the expectation value on the right-hand side of \eref{e:Cor} is taken, the amount of correlations between the two regions $S_X$ and $S_Y$.

Combining all the above results we obtain
\begin{eqnarray}\label{e:general_final}
\frac{\left|\langle A(t)B(t)\rangle_\mathrm{c}\right|}{\| A\|\| B\|}
&\leq& \mathscr{B}_r(t):=\Cor(S_X(r):S_Y(r))\nonumber\\
&&+4G(t)\left(\sum_{i\in X}\sum_{j\in S_X^\mathrm{c}(r)}+\sum_{i\in Y}\sum_{j\in S_Y^\mathrm{c}(r)}\right)F(\dist(i,j))
\end{eqnarray}
with
\begin{equation}\label{e:G}
G(t):=\frac{C+\| F\|}{C}\|\Phi\|\int_0^{|t|}\dd\tau g(\tau),
\end{equation}
where the integration of $g$ [as defined in \eref{e:g}] is elementary.

At this point we have made explicit the dependence of the right-hand side of \eref{e:general_final} on the radius $r<\dist(X,Y)/2$ of the balls $S_X$ and $S_Y$ \footnote{In principle, different radii could be chosen for $S_X$ and $S_Y$.}. Choosing $r$ small will in general reduce the contribution $\Cor(S_X(r):S_Y(r))$ stemming from the initial correlations, but will lead to a larger contribution from the Lieb-Robinson term, and {\em vice versa}. It is the interplay of these two contributions that can lead to interesting propagation patterns going beyond those that emerge from product initial states. An optimized bound $\mathscr{B}$ can be obtained by considering $r$ to be $t$-dependent, and minimizing the right-hand side of \eref{e:general_final} over $r(t)$ separately for each time $t$,
\begin{equation}\label{e:BoundFinal}
\frac{\left|\langle A(t)B(t)\rangle_\mathrm{c}\right|}{\| A\|\| B\|}
\leq\mathscr{B}(t):=\min_{0\leq r(t)\leq\dist(X,Y)/2}\mathscr{B}_{r(t)}(t).
\end{equation}
This amounts, at any fixed $t$, to making $r(t)$ just large enough to encompass the causal region to which the propagation is essentially restricted, but no larger, in order to reduce the contribution from correlations of the initial state. Initial correlations between certain regions become relevant only once those regions have been ``reached'' by the quasilocal dynamics. Equation \eref{e:BoundFinal}, along with \eref{e:Cor}--\eref{e:G}, is the main result of this paper.

\section{Single-site observables and specific types of interactions}
\label{s:specific}

A better intuition of the implications of the bound \eref{e:general_final} can be obtained by specializing the result to correlations between single-site observables (e.g., Pauli operators $\sigma_i$ in case of a spin-$1/2$ lattice model), and to specific types of interactions (e.g., nearest-neighbour or power law decaying interactions). 

Assuming single-site observables $A_i$ and $B_j$ supported at lattice sites $i\neq j$, \eref{e:general_final} simplifies to 
\begin{eqnarray}\label{e:single_final}
\frac{\left|\langle A(t)B(t)\rangle_\mathrm{c}\right|}{\| A\|\| B\|}
&\leq& \Cor(S_i(r):S_j(r))\nonumber\\
&&+4G(t)\left(\sum_{k\in S_i^\mathrm{c}(r)}F(\dist(i,k))+\sum_{k\in S_j^\mathrm{c}(r)}F(\dist(j,k))\right).
\end{eqnarray}
Estimates of the remaining summations in \eref{e:single_final} can be obtained by integral approximation. The bounds in the remainder of this section are less tight due to these further approximations, but their functional form becomes more evident.

\subsection{Finite-range or exponentially decaying interactions}
\label{s:finiterange}

As mentioned in section \ref{s:LR}, for interactions of finite range (like nearest-neighbour interactions) or for exponentially decaying interactions, the function $F(x)\propto\ee^{-ax}/(1+x)^{D+1}$ with $a>0$ is a suitable choice satisfying \eref{e:integrability} and \eref{e:reproducibility}. By integral approximation we can then bound
\begin{equation}
\sum_{k\in S_i^\mathrm{c}(r)}F(\dist(i,k))
\propto \!\!\!\sum_{k\in \Lambda:\dist(i,k)\geq r}\frac{\ee^{-a\dist(i,k)}}{(1+\dist(i,k))^{D+1}}\leq \frac{c\,\ee^{-ar}}{r^2}
\end{equation}
with a $D$-dependent constant $c>0$. Simplifying also the time-dependence in \eref{e:single_final} by estimating
\begin{equation}\label{e:Gsimple}
G(t) = \frac{C+\| F\|}{C}\|\Phi\|\int_0^{|t|}\dd\tau \left(\ee^{2\|\Phi\| C\tau}-1\right) \leq \frac{C+\| F\|}{2C^2}\ee^{2\|\Phi\| C|t|},
\end{equation}
we obtain 
\begin{equation}\label{e:exponential_final}
\frac{\left|\langle A(t)B(t)\rangle_\mathrm{c}\right|}{\| A\|\| B\|}
\leq \Cor(S_i(r):S_j(r)) + \frac{4c(C+\| F\|)\ee^{2\|\Phi\| C|t|-ar}}{C^2 r^2}.
\end{equation}
For the case of an uncorrelated initial state, $r=\dist(i,j)/2$ is the optimal choice for minimizing the second term on the left-hand side of \eref{e:exponential_final}, and one can further estimate
\begin{equation}\label{e:exponential_uncorrelated}
\frac{\left|\langle A(t)B(t)\rangle_\mathrm{c}\right|}{\| A\|\| B\|}
\leq \ee^{a(v|t|-\dist(i,j))}\times\mathrm{const.},
\end{equation}
similar to equation (3.1) of \cite{NachtergaeleOgataSims06}. An analogous result, only with different constants $v$ and $a$, holds also for initial states with exponentially clustered correlations \cite{BravyiHastingsVerstraete06}. In the presence of longer-ranged initial correlations, however, one would need to minimize \eref{e:exponential_final} over $r$ in order to obtain a tighter bound.

\subsection{Power law decaying interactions}

For power law decaying interactions we can choose $F(x)\propto (1+x)^{-\alpha}$ with $\alpha>0$. By integral approximation one can then bound
\begin{equation}\label{e:sum_power}
\sum_{k\in S_i^\mathrm{c}(r)}F(\dist(i,k))
\propto \!\!\!\sum_{k\in \Lambda:\dist(i,k)\geq r}\frac{1}{(1+\dist(i,k))^\alpha}\leq \frac{c}{r^{\alpha-D}}
\end{equation}
with a $D$-dependent constant $c>0$. Inserting \eref{e:Gsimple} and \eref{e:sum_power} into \eref{e:single_final} we obtain
\begin{equation}\label{e:powerlaw_final}
\frac{\left|\langle A(t)B(t)\rangle_\mathrm{c}\right|}{\| A\|\| B\|}
\leq \Cor(S_i(r):S_j(r))+\frac{4c(C+\| F\|)\ee^{2\|\Phi\| C|t|}}{C^2 r^{\alpha-D}}.
\end{equation}
As in section \ref{s:finiterange}, $r=\dist(i,j)/2$ is the optimal choice for minimizing the left-hand side of \eref{e:powerlaw_final} in the case of an uncorrelated initial state, yielding
\begin{equation}\label{e:powerlaw_uncorrelated}
\frac{\left|\langle A(t)B(t)\rangle_\mathrm{c}\right|}{\| A\|\| B\|}
\leq\frac{\ee^{v|t|}}{\dist(i,j)^{\alpha-D}}\times\mathrm{const.}
\end{equation}

For power law interactions with $\alpha>2D$ there exists a sharper Lieb-Robinson-type bound due to Foss-Feig \etal \cite{FossFeigGongClarkGorshkov15}, which could be used to derive a bound on equal-time correlation functions along the same lines as above.

\section{Examples of bounds for long-distance correlated initial states}
\label{s:examples}

One-dimensional examples are presented, illustrating how correlations of the initial state modify the creation and propagation of correlations between initially uncorrelated lattice sites.

\subsection{Connecting with an entangled pair}

Here we consider an initial state 
\begin{equation}\label{e:EntIni2}
|\psi\rangle=\left(|\uparrow\rangle_{-k}|\downarrow\rangle_k+|\downarrow\rangle_{-k}|\uparrow\rangle_k\right)\bigotimes_{m\neq -k,k}|\uparrow\rangle_m
\end{equation}
that is mostly of product form, except for a maximally entangled pair at sites $-k$ and $k$. We are interested in the time evolution of the connected correlation function $\langle\sigma_{-i}^z\sigma_i^z\rangle_\mathrm{c}$ between lattice sites $-i$ and $i$ that are a distance $\delta=2i$ apart. For the initial state \eref{e:EntIni2} we have
\begin{equation}\label{e:Cor_block}
\Cor(S_{-i}(r):S_i(r)) = \langle\phi_0|\sigma_{-k}^z\sigma_k^z|\phi_0\rangle_\mathrm{c}\,\Theta(r-|i-k|) = \Theta(r-|i-k|)
\end{equation}
for $r\leq i$, where $\Theta$ denotes the Heaviside step function. Assuming a lattice model with nearest-neighbour interactions, we combine \eref{e:exponential_final}, \eref{e:exponential_uncorrelated}, and \eref{e:Cor_block} to obtain
\begin{eqnarray}\label{e:bound_block}
\left|\langle\sigma_{-i}^z(t)\sigma_i^z(t)\rangle_\mathrm{c}\right| &\leq& \min\left\{1,\min_r\left(\Theta(r-|i-k|)+\tilde{c}\,\ee^{a(v|t|-r)}\right)\right\}\nonumber\\
&=&\min\left\{1,\tilde{c}\,\ee^{a(v|t|-|i-k|)}\right\},
\end{eqnarray}
where we have also included the bound
\begin{equation}
\left|\langle\sigma_{-i}^z(t)\sigma_i^z(t)\rangle_\mathrm{c}\right|\leq1,
\end{equation}
\begin{figure}\centering
\includegraphics[height=0.39\linewidth]{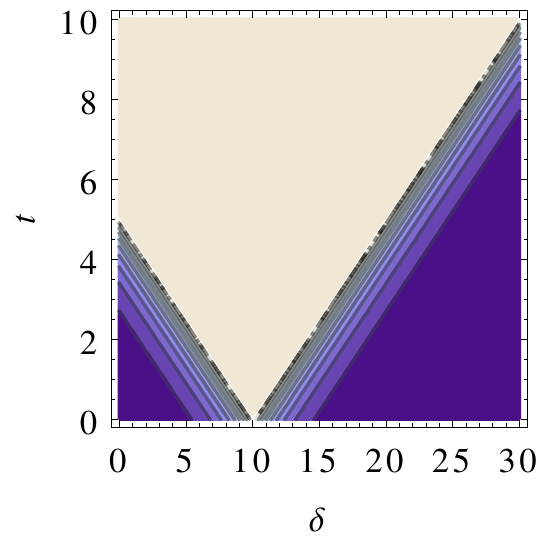}
\includegraphics[height=0.39\linewidth]{./LegendContour.pdf}
\includegraphics[height=0.39\linewidth]{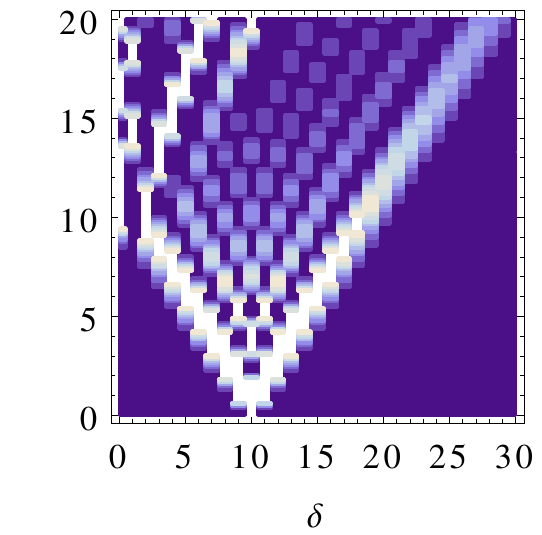}
\includegraphics[height=0.39\linewidth]{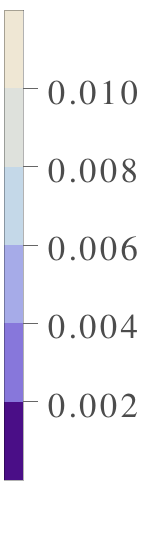}
\caption{\label{f:pair}%
Left: Contour plot of the bound \eref{e:bound_block} for the initial state \eref{e:EntIni2} and parameter values $2k=10$ and $a=v=\tilde{c}=1$. Right: Exact analytic results for the absolute value of equal-time correlations in an $XX$-chain with nearest-neighbour interactions \eref{e:XX}, starting from the initial state \eref{e:EntIni2}. In both plots, cone-like spreading of correlations similar to the behaviour in the absence of initial correlations is observed, but with an offset (i.e., a shift to higher values of $\delta$) by $2k$.
}
\end{figure}%
which follows from the Cauchy-Schwarz inequality. The contour plots in figure \ref{f:pair} (left) illustrate the creation of correlations in time and as a function of the distance $\delta$ between spins. The correlated pair leads to an effective reduction of the distance $\dist(i,j)$ between the spins by the distance $2k$ between the correlated sites: Mediated by the entanglement of the initial state, spins that are $2k+1$ sites apart ``feel'' each other as if they were neighbours. Accordingly, the time to transmit a signal across such an entangled quantum channel is reduced by $2k/v$, where $v$ is the Lieb-Robinson velocity occurring in \eref{e:bound_block}.

To assess how well the bound compares to the actual dynamics, we investigate the time-evolution of the state \eref{e:EntIni2} under an 
$XX$-Hamiltonian with nearest-neighbour interactions,
\begin{equation}\label{e:XX}
H=-J\sum_i \left(\sigma_i^x \sigma_{i+1}^x + \sigma_i^y \sigma_{i+1}^y\right),
\end{equation}
where we set $J=1$. 
As shown in figure \ref{f:pair} (right), correlations spread in the interior of a cone, with a spatial offset of $2k=10$ compared to the case of an uncorrelated initial state. This confirms that the bound \eref{e:bound_block} reproduces the qualitative features of the dynamics, although, as expected, the velocity at which magnon quasiparticles propagate in the $XX$-chain is slower than the estimated velocity $v$ occurring in the estimate \eref{e:bound_block}. The bound and the model calculation shown in figure \ref{f:pair} can be seen as simplified illustrations of the enhancement of quantum transport and qubit transfer in a long-distance entangled quantum channel, as mentioned in item \ref{b} of the Introduction.

\subsection{Power law clustering of initial correlations}
\label{s:powerlaw}

\begin{figure}\centering
\includegraphics[width=\linewidth]{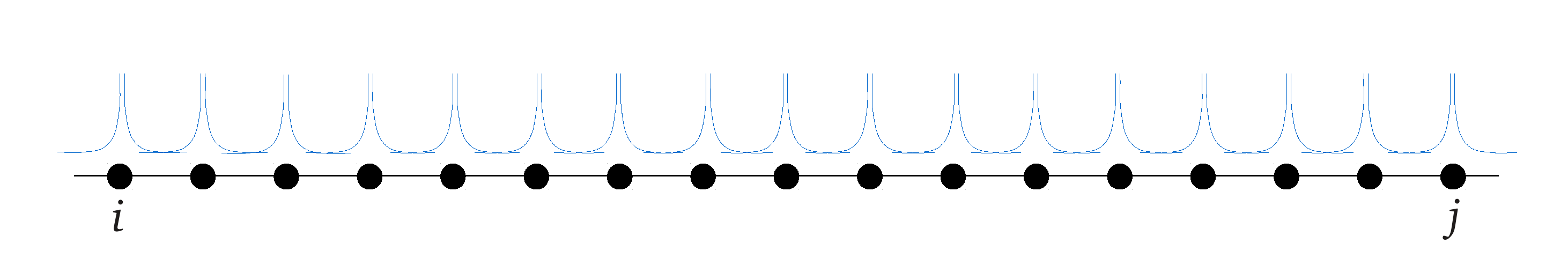}
\includegraphics[height=0.39\linewidth]{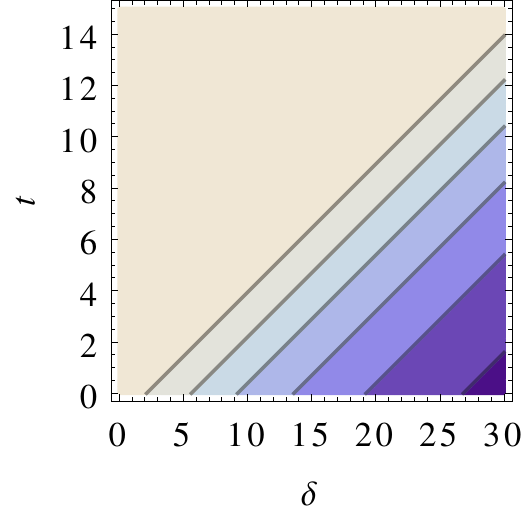}
\includegraphics[height=0.39\linewidth]{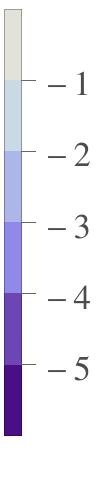}
\includegraphics[height=0.39\linewidth]{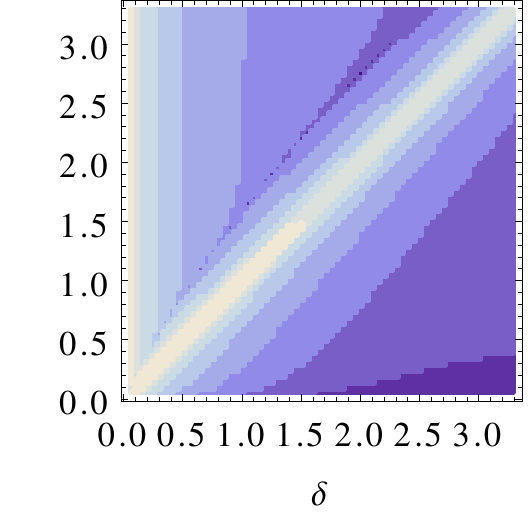}
\includegraphics[height=0.39\linewidth]{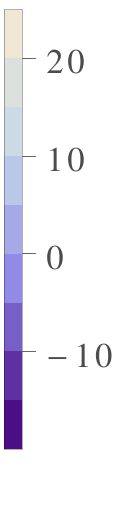}
\includegraphics[width=0.49\linewidth]{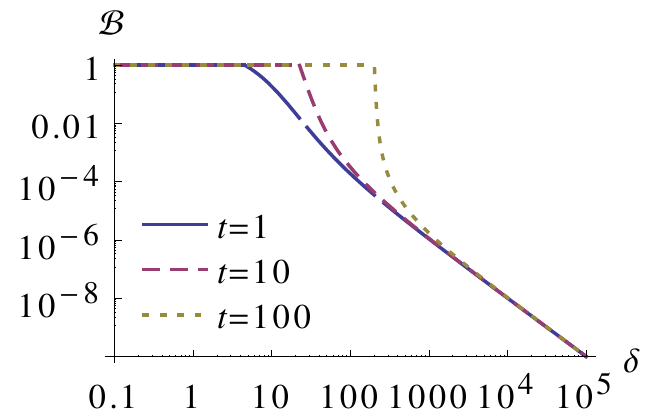}
\includegraphics[width=0.49\linewidth]{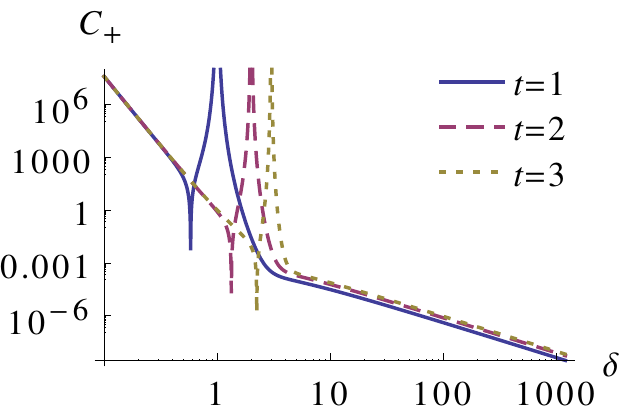}
\caption{\label{f:power}%
Top: Sketch of a spin chain with power law clustering of correlations \eref{e:Cor_powerlaw}. Centre left: Contour plot of the logarithm of the bound \eref{e:bound_power} for power law-correlated initial states, with parameters $\chi=2$, $a=v=c_1=c_2=1$. Bottom left: As above, but for fixed values of $t$ and in a log-log representation. Enter right: Exact analytic results for the logarithm of the absolute value of the equal-time correlation function for a Kondo impurity coupled to a Fermi sea at zero temperature. Bottom right: As above, but for fixed values of $t$ and in a log-log representation. All these plots show that correlations spread inside a cone, where the spatial decay outside the cone follows a power law, as is visible in the log-log plots.
}
\end{figure}%

Items \ref{a} and \ref{c} of the Introduction describe two possible scenarios of physical interest where initial states with power law-clustered correlations arise (illustrated in figure \ref{f:power} top). In that case the connected correlations initially satisfy
\begin{equation}\label{e:Cor_powerlaw}
\left|\langle\sigma_i^z\sigma_j^z\rangle_\mathrm{c}\right|\leq\frac{c_1}{\dist(i,j)^\chi}
\end{equation}
with some exponent $\chi\geq0$. Assuming again a chain of spin-$1/2$ degrees of freedom with nearest-neighbour interactions, we combine \eref{e:exponential_final}, \eref{e:exponential_uncorrelated}, and \eref{e:Cor_powerlaw} to obtain the bound
\begin{equation}\label{e:bound_power}
\left|\langle\sigma_i^z(t)\sigma_j^z(t)\rangle_\mathrm{c}\right|\leq\min\left\{1,\min_r\left(\frac{c_1}{(\dist(i,j)-2r)^\chi}+c_2\,\ee^{a(v|t|-r)}\right)\right\}.
\end{equation}
As illustrated in figure \ref{f:power} (left), this bound shows cone-like propagation, but, in contrast to the case without initial correlations, the spatial decay outside the cone follows a power law with exponent $\chi$ (instead of an exponential decay).

As a physical illustration of this kind of propagation behaviour, we borrow results from Medvedyeva \etal \cite{MedvedyevaHoffmannKehrein13} on the spatiotemporal build-up of the Kondo screening cloud. The authors of that paper study correlations between the spin of an impurity and the spin of a conduction electron in the three-dimensional Kondo model. Strictly speaking this model does not satisfy the conditions under which the Lieb-Robinson bound \eref{e:LR} has been proved, and hence our bound \eref{e:BoundFinal} does not apply. Proving Lieb-Robinson bounds for general bosonic or fermionic hopping models turns out to be elusive, but model calculations indicate that the majority of such models with local interactions nonetheless do show lightcone dynamics. Counterexamples exist, but they require careful design \cite{EisertGross09}, and we have good reason to believe that the Kondo model can well serve as an example illustrating the physics described by the bound \eref{e:BoundFinal}.

The initial state $|\uparrow\rangle\otimes|\mathrm{FS}\rangle$ used in \cite{MedvedyevaHoffmannKehrein13} is a product of the impurity spin and the Fermi sea of the conduction electrons. While the impurity spin is initially uncorrelated with the conduction electrons, the conduction electrons themselves are spatially correlated among one another. At zero temperature these initial correlations decay like a power law in space. Analytic expressions are then obtained for the correlation functions of the Kondo model at the Toulouse point (i.e., for a special value of one of the coupling constants in the Kondo Hamiltonian), see equations (9) and (19)--(23) of \cite{MedvedyevaHoffmannKehrein13}. Numerically evaluating these equations in the zero-temperature limit, we obtain the spatiotemporal spreading of the equal-time correlation functions plotted in figure \ref{f:power} (right). The correlations are sharply peaked on the boundary of the lightcone. The spatial decay outside the cone follows a power law, in 
agreement with the 
bound shown in figure \ref{f:power} (left).

\section{Conclusions}

In this paper we have developed theoretical tools, applicable to a broad class of quantum lattice models, for the description of the propagation of correlations in the presence of initial entanglement. The main result \eref{e:general_final}--\eref{e:BoundFinal} is an upper bound on the connected correlation function, containing a Lieb-Robinson-type contribution capturing the dynamics, and a second term that takes into account initial correlations. Depending on the amount and shape of initial correlations, the bound is able to capture the extremes of product initial states (as in figure \ref{f:XX_Corr} left) on the one side and long-distance entangled initial states (as in figure \ref{f:XX_Corr} right) on the other side, as well as the more involved cases in between where the interplay of dynamics and initial entanglement leads to nontrivial propagation patterns.

The essential prerequisite for the proof is that some kind of Lieb-Robinson bound can be derived for the system under consideration. We have here used the setting of \cite{NachtergaeleOgataSims06} (detailed in section \ref{s:LR}), which includes short- as well as long-range interacting lattice models, but other settings may be used to either allow for different or more general types of interactions, or for sharper bounds \cite{NachtergaeleSims10,KlieschGogolinEisert14,Damanik_etal14,StorchvandenWormKastner15}. The main idea of the proof is to divide the lattice into two regions, one consisting of two disjoint spheres $S_X$ and $S_Y$ of radius $r$, centred around the lattice sites $i$ and $j$ for which the connected correlation function $\langle\sigma_i^x\sigma_j^x\rangle_\mathrm{c}$ is to be estimated; and the other region being the complement of the two spheres. The size of the correlations can then be bounded by two contributions: The first two terms on the right-hand side of \eref{e:split} account for 
the propagation of correlations due to interactions, and they decrease with increasing $r$; the third term on the right-hand side of \eref{e:split} accounts for initial correlations between $S_X$ and $S_Y$, and it increases with $r$. An optimal bound is then found by choosing, for each time $t$ and given sites $i$ and $j$, the optimal value of $r$ for which the bound becomes minimal, as in the final result \eref{e:BoundFinal}.

Owing to its generality, many potential applications of this result can be envisaged. Physical applications in which entangled initial states have a strong effect on the propagation of correlations include quenches away from a quantum critical point, quantum transport and qubit transfer in spin chains with long-distance entangled ground states, or the building-up of a Kondo screening cloud at zero temperature. Some of these examples, or simplified toy models of them, have been discussed in section \ref{s:examples}. While absolute magnitudes and propagation velocities are overestimated (as is generally the case when using Lieb-Robinson bounds), the bounds on correlation functions derived in this paper show good qualitative agreement with model calculations for entangled initial states. These comparisons were made for integrable models where exact analytic results are available, but we have no reason to believe that the propagation of correlations will in general be substantially different in nonintegrable 
systems. In certain instances, however, localisation effects can strongly suppress the propagation of correlations. In these cases the bound \eref{e:BoundFinal} should be based on a sub-ballistic Lieb-Robinson bound like the one derived for disordered spin systems in \cite{BurrellOsborne07}.

We believe that the results here provide useful descriptions of the propagation patterns to be expected in a variety of physical situations of interest, and they should be particularly expedient for larger system sizes where numerical simulations are out of 
reach.\\

\ack
\addcontentsline{toc}{section}{Acknowledgments}
Helpful communications with Fabian Essler, Alexey Gorshkov, and Stefan Kehrein are gratefully acknowledged.
The author is financially supported by the National Research Foundation of South Africa through the Incentive Funding and the Competitive Programme for Rated Researchers.


\section*{References}
\addcontentsline{toc}{section}{References}
\bibliographystyle{iopart-num}
\bibliography{LRLR}

\end{document}